\documentclass[paper,notoc,nohyper]{JHEP} 
\usepackage{epsfig}
\usepackage{amsmath}
\usepackage{amssymb}
\usepackage{cite}

\def\cF{{\cal{F}}}
\def\cO#1{{\cal O}\left( {#1} \right)}
\def\asb{{\bar \alpha}_{\mbox{\scriptsize s}}}
\def\as{\alpha_{\mbox{\scriptsize s}}}
\def\ga{\gamma}
\def\om{\omega}

\def\ca{C_A}

\def\nf{n_f}

\def\half{\mbox{\small $\frac{1}{2}$}}

\def\ho{\half\om}
\def\bd{{\bar d}}

\title{A resummation of large sub-leading corrections at small
  $\boldsymbol{x}$.}

\author{G.P. Salam\thanks{Work supported by E.U. QCDNET contract
    FMRX-CT98-0194.}  \\
  INFN, Sezione di
  Milano, Via Celoria 16, Milano, Italy\\
  E-mail: \email{G.Salam@mi.infn.it}}
  
\abstract{The NLL corrections to the BFKL kernel are known to be very
  large, to the extent that even for small values of $\as$, they lead
  to physical cross sections which are not positive definite. It is
  shown in the context of a toy model, that such pathological
  behaviour is an artifact of the truncation at NLL order, and is
  associated in particular with double transverse logarithms. These
  are resummed in a manner consistent with the full NLL kernel, and
  are shown to change its properties quite considerably.}

\preprint{hep-ph/9806482\\
          IFUM-627-FT\\
          June 1998}


\begin{document}

\section{Introduction}
\label{sec:intro}

Following many years of calculations \cite{NLL} the full NLL
corrections to the BFKL kernel \cite{BFKL} recently became available
\cite{FL,CC98}. Quite soon afterwards it was pointed out that there
were serious problems with the convergence of the kernel \cite{BV},
and in particular that for values of $\as$ above about $0.05$ the
saddle-point structure of the characteristic function changes: instead
of having a single saddle-point on the real $\gamma$ axis, one has two
complex-conjugate saddle points \cite{Ross}. As a result the power of
the small-$x$ growth, rather than depending on $\as$ (schematically)
as $\asb 4\ln2 - N\asb^2$, where $N$ is a number, seems to acquire a
more complex (non-parabolic) dependence on $\as$. A particularly
worrying consequence of the saddle points having complex values is
that the solutions oscillate as a function of transverse momentum
(this is a statement which is valid beyond the saddle-point
approximation), and even when integrated with the appropriate
matrix-elements can lead to negative results for physical cross
sections \cite{Levin}.

This paper will consider this pathological behaviour in the context of
yet higher order corrections. The discussion will be fundamentally
related to the issue of the choice of scales (discussed also in some
detail in \cite{CC98}). In determining the leading contribution to the
cross section for the high-energy scattering of two objects with
transverse scales $k_1^2$ and $k_2^2$, one resums terms of the form
\begin{equation*}
  \left(\as \ln \frac{s}{s_0} \right)^n.
\end{equation*}
At leading-logarithmic (LL) (in $x$) order, the choice of $s_0$ is
immaterial. When one starts to consider next-to-leading-logarithmic
(NLL) terms,
\begin{equation*}
  \as \left(\as \ln \frac{s}{s_0} \right)^n,
\end{equation*}
one sees that their form will depend on the choice made for $s_0$ in
the LL terms. Various choices come to mind for $s_0$. A symmetrical
$s_0=k_1k_2$ is a Regge-motivated choice. But in situations where
$k_1^2 \gg k_2^2$ (or $k_2^2 \gg k_1^2$) it is more natural to take
$s_0 = k_1^2$ ($s_0 = k_2^2$), because that is the scale which enters
in the appropriate DGLAP-type resummation \cite{DGLAP}. In this limit
the cross section has terms of the form
\begin{equation*}
  \left(\as \ln \frac{k_1^2}{k_2^2} \,\ln \frac{s}{k_1^2} \right)^n.
\end{equation*}
One sees immediately that trying to rewrite them in terms of
$s_0=k_1k_2$ leads to double transverse logarithms in
the series,
\begin{equation*}
   \left(\as \ln^2 \frac{k_1^2}{k_2^2}\right)^{n-m}  
     \left(\as \ln \frac{k_1^2}{k_2^2} \,\ln \frac{s}{k_1k_2}\right)^m\,.
\end{equation*}
Though formally subleading, these terms form a significant part of the
NLL corrections to the BFKL kernel, because the double logarithm can
be large even when $|\ln k_1^2 /k_2^2| \ll \ln s/s_0 $.

To study such effects in more detail it is convenient to go the
Mellin-transform space of the gluon's Green function (assuming
straightforward exponentiation, in the
notation of \cite{CC98}),
\begin{equation}
  \label{eq:eigen}
    \int \frac{d\ga}{2\pi i}\, \frac{d\om}{2\pi i}\,
    \left(\frac{s}{s_0}\right)^{\omega} g_\om(\ga)\;
    \frac{1}{k_1^2} \! \left( \frac{k_1^2}{k_2^2} \right)^{\!\ga}
    \,, \quad 
    g_\om(\ga) = \frac{F(\ga)}{1 - \frac{\asb}{\om} \chi(\ga)},
\end{equation}
with $\asb = \as\ca/\pi$, $F(\ga)$ a combination of $\om$-independent
kernels, and $\chi(\ga)$ known as the characteristic function. At LL
order \cite{BFKL},
\begin{equation}
  \label{eq:chill}
  \chi(\ga) = 2\psi(1) - \psi(\ga) - \psi(1-\ga)\,.
\end{equation}
The effect of the changing $s_0$ from $k_1^2$ to $k_1k_2$ is to take
$\ga$ to $\ga+\half \omega$. As has been discussed by both Fadin and
Lipatov \cite{FL} and Ciafaloni and Camici \cite{CC98}, such a
transformation, acting on the LL kernel, produces a term
\begin{equation*}
  -\frac{\asb}{2\ga^3}\,,
\end{equation*}
which is precisely as found in the NLL kernel (with scale $s_0=k_1
k_2$). It turns out that this, and the corresponding $\asb/2(1-\ga)^3$
term (which are the first of the double-logarithmic terms discussed
above), are responsible for about half the NLL correction to the
asymptotic BFKL exponent, $\asb\chi(\half)$.

The purpose of this article is to resum the double logarithmic terms.
The basis of the discussion will be that with a scale $s_0=k_1^2
\,(k_2^2)$ the appropriate DGLAP limit $\ga\to0$ ($1-\ga\to0$) must be
free of double logarithms, i.e.\ there should be no terms $\as^n
\ga^{-k}$ ($\as^n (1-\ga)^{-k}$) with $k>n+1$.

In section 2, we will consider a toy kernel, a variant of the
linked-dipole-chain (LDC) model \cite{LDC}, which will have the
feature of reproducing the BFKL kernel at LL order, and of resumming
all the ``leading'' double-logarithmic (DL) terms, $\as^n/\ga^{2n+1}$.
Its main interest will be related to the study of the general analytic
structure and the order-by-order (non-)convergence of the kernel. It
will be shown that the feature noted by Ross \cite{Ross} at NLL order,
namely that the LL saddle point splits into two symmetrical saddle
points which no longer lie on the real $\ga$ axis, is reproduced by
this model. This happens at a value of $\as$ considerably smaller than
the point at which the NLL corrections cause the asymptotic exponent
to become negative --- essentially the NLL corrections have a much
larger effect on the second derivative, $\chi''(\half)$ than on
$\chi(\half)$ itself.

If one resums the toy kernel, one returns to a single saddle point on
the real axis. This is reassuring since the presence of saddle-points
off the real axis would almost certainly have led to oscillating
physical cross sections \cite{Ross,Levin}. It is to be noted that for
reasonable values of $\as$ ($>0.1$) it is {\emph not} sufficient
simply to go to NNLL or in general N$^n$LL order to obtain a
well-behaved kernel: the pathologies seen in the NLL kernel persist
for any fixed $n$, hence the resummation of an appropriate subset of
the N$^n$LL-order terms, for all $n$, is essential.

Section 3 will discuss a way, given a fixed order kernel up to order
$\as^N$, to resum terms
\begin{equation*}
  \frac{\as^n}{\ga^k}\,,\; \frac{\as^n}{(1-\ga)^k} \qquad  2n+1-N \le
  k \le 2n+1\,. 
\end{equation*}
The technique will be applied to the NLL kernel (with details given in
the appendix). The procedure is not unique, in that it introduces
subleading terms, with $n>N$, $k<2n+1-N$ which are not under control.
The study of four schemes which differ in their treatment of these
subleading terms suggests that the value of $\chi(\half)$ is
reasonably stable, i.e.\ scheme-independent after the resummation. On
the other hand $\chi''(\half)$ remains quite unstable, being sensitive
in particular to one's treatment of terms $\as^n/\gamma^{n+1}$.

\section{An LDC-like toy kernel}
\label{sec:ldc}

The question of how to construct a small-$x$ kernel which reproduces
the DGLAP double-logarithmic ($\ln x \ln Q^2$) limit both for $k_1\gg
k_2$ and for $k_1 \ll k_2$ was considered a few years ago by
Andersson, Gustafson and Samuelsson \cite{LDC}. One of the equations
that they proposed for the unintegrated gluon distribution $\cF$
was\footnote{Note that the equation describing the actual LDC model,
  as used for example in the LDC Monte Carlo event generator
  \cite{LDCmc}, differs from BFKL at LL order, as discussed in
  \cite{LDC}.}
\begin{equation}
  \label{eq:LDCbfkl}
  \frac{d \cF(x,k^2)}{d \ln (1/x)} = \frac{\asb}{\pi}
  \int \frac{d^2q}{\pi q^2} \left( \cF(x',|k+q|^2) -
    \cF(x,k)\,\Theta(k-q)\right) \,.
\end{equation}
Here, in $\cF(x,k_1^2)$, $x$ is defined as $k_1^2/s$, so that we are
using $s_0 = k_1^2$.  The difference between this and the BFKL
equation lies in the presence of $x' = \max(x,x |k+q|^2/k^2)$ in the
right hand side. The justification is that in a situation where
$|k+q|^2\gg k^2$, from the DGLAP point of view, $x$ is no longer the
appropriate evolution variable, but rather (within the
double-logarithmic approximation) $x\cdot|k+q|^2/k^2$.

Taking the Mellin transform one sees that the characteristic function
satisfies the equation (where for now we stay with the scale
$s_0=k_1^2$)
\begin{equation}
  \label{eq:LDCchi}
  \chi(\ga) = \frac\omega\asb = 2\psi(1) - \psi(\ga) -
  \psi(1-\ga+\omega)\,.
\end{equation}
In the limit $\asb\to0$ it explicitly reproduces the usual LL
characteristic function, \eqref{eq:chill}. In the DGLAP limit,
$k_1^2\gg k_2^2$, corresponding to the region close to $\ga=0$, one
has only the pole $1/\gamma$ --- there are no double logarithms, as is
required with the scale choice $s_0=k_1^2$. In the opposite DGLAP
limit, $k_2^2\gg k_1^2$, the relevant region of $\ga$ is close to
$\ga=1$. Making the transformation to the relevant scale, $s_0=k_2^2$,
using $\ga\to\ga+\omega$, there is only the pole $1/(1-\ga)$. So one
is completely free of double transverse logarithms in both DGLAP
limits.

In practice it will be most convenient to work in terms of the
symmetric scale choice, $s_0=k_1k_2$, with which \eqref{eq:LDCchi}
becomes
\begin{equation}
  \label{eq:chiimp}
  \chi(\ga) = \frac\omega\asb = 2\psi(1) - \psi(\ga + \half\omega) -
  \psi(1-\ga+\half\omega)\,.
\end{equation}
Iterating, it is straightforward to obtain the higher order
corrections, $\chi_n$, (note the difference in normalisation compared
to \cite{FL,CC98})
\begin{equation*}
  \chi(\gamma) = \sum_{n=0} \asb^n \chi_n(\gamma)\,.
\end{equation*}
The first two are 
\begin{align}
  \label{eq:chi1}
  \chi_1 &= -\frac12 \chi_0(\gamma)\, [\psi'(\gamma) +
  \psi'(1-\gamma)]\,,\\
  \label{eq:chi2}
  \chi_2 &= -\frac12 \chi_1(\gamma) \, [\psi'(\gamma) +
  \psi'(1-\gamma)] - \frac18 \chi_0(\gamma)^2\, [\psi''(\gamma) +
  \psi''(1-\gamma)]
\,.
\end{align}
While $\chi_1$ does not correspond exactly to any particular piece of
the full NLL kernel, it does reproduce the most divergent
($1/\gamma^3$) part of the kernel in the regions of $\gamma$ close to
$0$ and close to $1$:
\begin{subequations}
  \label{eq:chi1exp}
\begin{align}
  |\gamma| \ll 1:& \quad \chi_1=-\frac{1}{2\gamma^3} 
  -\frac{\pi^2}{6\gamma} - \zeta(3) +  \cO{\gamma}\,,\\
  |1-\gamma| \ll 1:& \quad \chi_1=-\frac{1}{2(1-\gamma)^3} 
  -\frac{\pi^2}{6(1-\gamma)} -\zeta(3) + \cO{1-\gamma}\,.
\end{align}
\end{subequations}
In fact, up to $\cO{1/\gamma}$ it has the same expansion as the
following part of the kernel (in the notation of Fadin and Lipatov
\cite{FL}):
\begin{equation*}
  -4\left(4\phi(\ga) - \psi''(\ga) - \psi''(1-\ga)\right)\,,
\end{equation*}
though this may well be of little significance.

\subsection{Double-logarithmic and resummed structure}

To study the leading double-logarithmic structure of the kernel it
suffices to examine the behaviour related just to the pole at $\ga=0$, 
i.e.\ a ``left-hand piece,'' $\chi_L$, of the characteristic function, 
\begin{equation}
  \label{eq:chiL}
  \chi_L(\ga) = \frac{1}{\ga + \half \asb\chi_L}\,.
\end{equation}
The remaining constant piece (including the part coming from
$\psi(1-\ga+\half\omega)$) has no effect on the leading $\ga=0$ DL
structure. The solution of \eqref{eq:chiL} is
\begin{equation}
  \label{eq:chiLsln}
  \chi_L(\gamma) = \frac{-\gamma \pm \sqrt{\gamma^2 + 2\asb}}{\asb}\,.
\end{equation}
It has different expansions according to the sign of $\gamma$:
\begin{subequations}
\begin{align}
  \label{eq:chiLser}
  \gamma > 0 &: \quad \chi_L = \frac{1}{\gamma} + 
                \sum_{n=1}^\infty d_{n,2n+1}\,
                \frac{\asb^n }{\gamma^{2n+1}}\,,\\
  \gamma < 0 &: \quad \chi_L = -\frac{2\gamma}{\asb} - \frac{1}{\gamma} - 
                \sum_{n=1}^\infty  d_{n,2n+1}\,
                \frac{\asb^n}{\gamma^{2n+1}}\,,
\end{align} 
\end{subequations}
where
\begin{equation}
  \label{eq:chiLsercoeffs}
  d_{n,2n+1} = (-1)^n \,\frac{(2n-1)!}{2^{n-1}\,(n-1)!\,(n+1)!}\,.
\end{equation}
Performing an expansion around $\ga=1$ one finds an analogous series
in $(1-\ga)$. It should be emphasised that the result
\eqref{eq:chiLser} for the leading double logarithmic structure is not
in any way model-dependent. It depends only on the fact that with the
scale $s_0=k_1^2$, the kernel should be free of double logarithms
around $\ga=0$, and that the leading (in $\as$) divergence there is
$1/\ga$.

\begin{figure}
  \begin{center}
    \epsfig{file=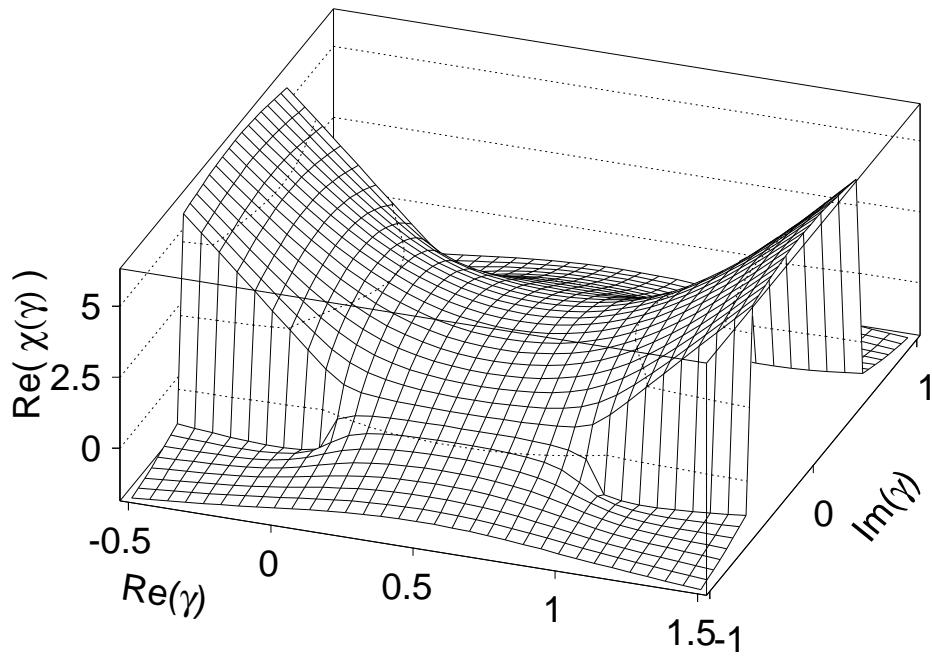,width=0.8\textwidth}
    \caption[]{The real part of the toy kernel in the complex plane,
      shown for $\asb=0.2$.}
    \label{fig:chicomp}
  \end{center}
\end{figure}

Potentially more model dependent is the behaviour in the regions $\ga$
and ($1-\ga$) very negative, where one sees\footnote{Analogous
  behaviour has also been observed \cite{future} in the context of the
  CCFM equation \cite{CCFM}, though the slope is different.} $\chi
\simeq -2\ga/\asb$ and $\chi \simeq -2(1-\ga)/\asb$ respectively.  The
solution of \eqref{eq:chiimp}, shown in figure~\ref{fig:chicomp}
illustrates this behaviour. The branch points correspond (at least
qualitatively) to those in \eqref{eq:chiLsln}. As $\asb$ goes to zero,
they move closer to $\ga=0$ and to $\ga=1$, while the sections in
between them grow steeper, increasing the all-order kernel's
similarity to the shape which is familiar in the LL limit (i.e.\ poles
at $\ga=0$ and $\ga=1$).

\subsection{Convergence of the toy kernel}
\label{sec:toyconv}

\begin{figure}
  \begin{center}
    \input{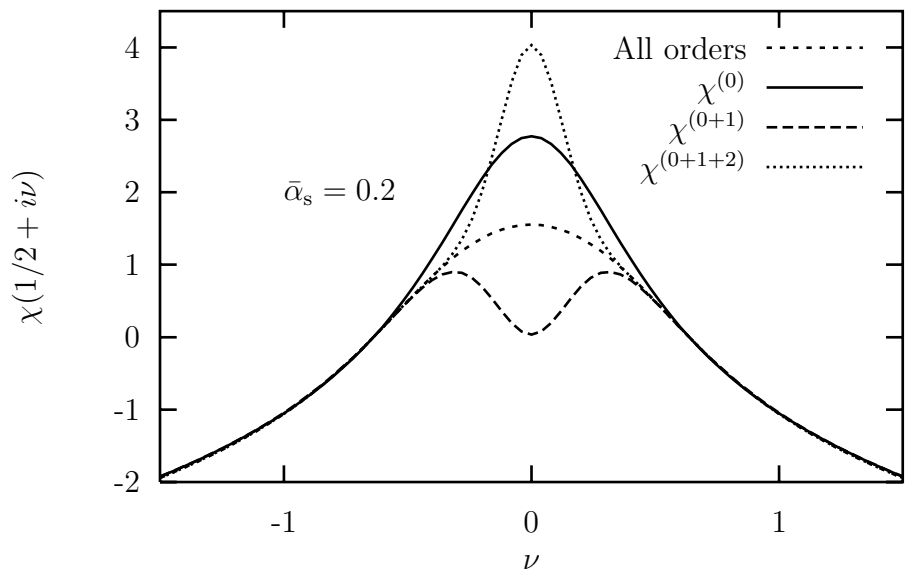}
    \caption[]{The toy kernel along the imaginary axis at various
      orders. Also shown is the resummed result.}
    \label{fig:saddles}
  \end{center}
\end{figure}

The toy kernel is interesting also from the point of view of its
convergence properties. Figure \ref{fig:saddles} shows the structure
of the kernel along the line $\gamma=1/2 + i\nu$ at various orders for
$\asb=0.2$.  As in the full NLL kernel \cite{Ross}, there are two
symmetric saddle-points at non-zero $\nu$ at NLL order. But if one
adds in the NNLL terms one sees that they disappear. At NNNLL (not
shown) they come back again. One finds in general that for even orders
in $\asb$ there is a single saddle-point at $\nu=0$, while for odd
orders in $\asb$ the kernel has two symmetric saddle-points at
non-zero $\nu$.  Figure \ref{fig:saddles} shows also the all-order
resummed result (a slice of figure~\ref{fig:chicomp}), i.e.\ the
direct solution of \eqref{eq:chiimp}: it has a single saddle point at
$\nu=0$.

This seems to indicate that the splitting into two of the saddle-point
in the NLL BFKL kernel, and the associated oscillating solutions are
very much artifacts of the truncation of the kernel at finite order
(though it turns out, at least in the toy model, that at NLL order the
value of $\chi$ at the two saddle-points is a somewhat better
approximation to the resummed $\chi(\half)$, than the NLL
$\chi(\half)$).

\begin{figure} 
  \begin{center}
    \input{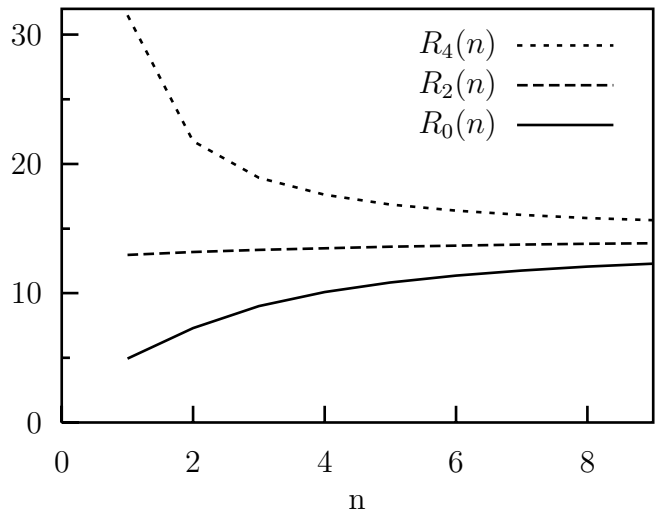}
    \caption[]{$R_0(n)$, $R_2(n)$, $R_4(n)$ as a function of the order
      of $\as$, $n$.}
    \label{fig:Rn}
  \end{center}
\end{figure}

To further study the poor convergence of the toy kernel, one can
examine the following ratios,
\begin{equation}
  \label{eq:Rpn}
  R_p(n) = -\left.
    \frac{d^p\chi_n/d\gamma^p} {d^p\chi_{n-1}/d\gamma^p}
    \right|_{\gamma=1/2}
    \,,
\end{equation}
as measures of the convergence of the expansion of $\chi$ and its
derivatives at $\gamma=\half$. They are plotted as a function of $n$ in
figure \ref{fig:Rn}.
At small orders, the convergence is far worse for the derivatives of
$\chi$ than for $\chi$ itself.  Hence at NLL order, $\chi''(\half)$
goes through zero (i.e.\ the saddle-point at $\nu=0$ splits into two)
long before $\chi(\half)$ itself does.  A simple explanation for the
relatively poor convergence of $\chi''(\half)$ compared to that of
$\chi(\half)$ is to be had by noting that if the NLL corrections are
dominated by a part
\begin{equation*}
  \chi_1 \simeq \frac1{\gamma^3} + \frac{1}{(1-\ga)^3},
\end{equation*}
then $\chi_1''(\half)/\chi_1(\half) = 48$, while a similar
approximation for the LL kernel gives a relation
$\chi_0''(\half)/\chi_0(\half) = 8$. The difference comes entirely
from the $3\times4$ obtained in differentiating $1/\gamma^3$ as
opposed to $1\times2$ in differentiating $1/\ga$.

\begin{figure}[htbp]
  \begin{center}
    \input{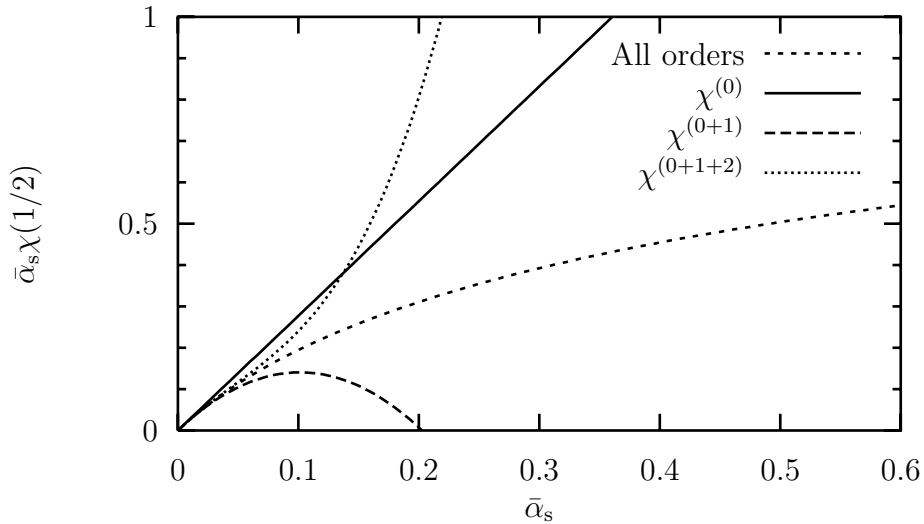}
    \caption[]{The toy-kernel asymptotic power as a function of $\asb$,
      at leading, NLL, NNLL and all orders.}
    \label{fig:chivasb}
  \end{center}
\end{figure}

Nevertheless, the perturbative expansion is rather poor even for
$\chi(1/2)$, as can be seen from figure~\ref{fig:chivasb} which shows
$\asb\chi(\half)$ as a function of $\asb$ at different orders. For
realistic values of $\asb$, say $0.2$, the fixed-order expansions are
quite unreliable --- and going from NLL to NNLL does not bring one any
closer to the all-order result except for $\as\lesssim 0.1$.

To summarise, what one learns from this toy kernel is that the double
transverse logarithms lead to a very poor convergence of the BFKL
kernel at subleading orders. The features at NLL order (the poor
convergence of $\chi(\half)$, the worse convergence of the second
derivative, leading to the splitting in two of the saddle-point)
are very similar to those observed in the full NLL kernel. But they
are quite unrelated to the characteristics (both the analytic
structure and the value of $\chi(\half)$) observed after a
resummation.  Thus the importance of a consistent resummation in
the case of the full NLL kernel.

\section{Improving the full NLL kernel}
\label{sec:ifn}

\subsection{A resummation procedure}
\label{sec:resum}

Schematically, the strategy for the construction of a kernel with the
appropriate set of double logarithms will consist in the replacement
of divergences $\as^n/\ga^k$ and $\as^n/(1-\ga)^k$ with terms
$\as^n/(\ga+\half\om)^k$ and $\as^n/(1-\ga+\half\om)^k$, while
maintaining the correct expansion to $\cO{\as^n}$.

To understand how to do so systematically, for a kernel known to
arbitrary order (the explicit results for NLL order are given in the
appendix), let us start with a discussion of the relation between the
pattern of divergences in a kernel with $s_0=k_1^2$, $\xi(\ga)$ and
one with $s_0=k_1k_2$, $\chi(\ga)$, which are related through
$\chi(\ga) = \xi(\ga+\ho)$ with $\om=\asb\chi(\ga)$. We will make the
assumption that $\xi(\ga)$ has an expansion of the form
\begin{equation*}
  \xi(\ga) = \sum_{n=0} \sum_{k=-\infty}^{n+1} C_{n,k} \frac{\asb^n}{
    \ga^k}\,
\end{equation*}
around $\ga=0$. The upper limit on $k$ embodies the statement that
$\xi(\ga)$ is free of double logarithms. If we substitute
$\ga\to\ga+\ho$ to obtain $\chi(\ga)$, we immediately see that
$\chi(\ga)$ will contain a set of terms $\asb^n \ga^{-2n+1}$ arising
from
\begin{equation*}
  \frac{1}{\ga} \longrightarrow \frac{1}{\ga + \frac{\asb}{2\ga}}\,,
\end{equation*}
(where we have put in $C_{0,1}=1$). These terms belong to the set
resummed in \eqref{eq:chiLser}. Substituting them into the most
divergent term of the NLL kernel, $\as/\ga^2$, 
$$
  C_{1,2} \,\frac{\asb}{\ga^2} \longrightarrow
  C_{1,2} \,\frac{\asb}{(\ga+\cO{\as^{n+1}/\ga^{2n+1}})^2}\,,
$$
and expanding, leads to a set of sub-leading DLs:
$$
\frac{\as}{\ga^2}\,\frac{\as^{n+1}}{\ga^{2(n+1)}}\,.
$$
together with less leading DLs.
More generally the substitution, 
\begin{equation*}
  C_{n,k} \,\frac{\asb^n}{\ga^k} \longrightarrow
  C_{n,k} \,\frac{\asb^n}{(\ga+\half\asb\chi)^k}
\end{equation*}
will lead to a set of terms in $\chi(\ga)$ of the form
\begin{equation*}
   \frac{\asb^n}{\ga^k} \left(\frac{\as}{\ga^2}\right)^{m-n} = 
   \frac{\asb^{m}}{\ga^{2(m-n) + k}}
\end{equation*}
for $m\ge n$, together with less divergent terms. So if one has the
perturbative series for $\xi(\ga)$ up to order $N$, in shifting to get
$\chi(\ga)$ one will unambiguously reproduce all double logarithms
which are more divergent than those produced by shifting the most
divergent unknown term in $\xi$, i.e.\ $\asb^{N+1}\ga^{-N-2}$, which
would give, in $\chi(\ga)$, terms of the form $\asb^n{\ga^{-(2n-N)}}$,
for $n>N$. In other words one correctly reproduces in $\chi(\ga)$ all
terms of $\cO{\as^n}$ for $n\le N$, and additionally
double-logarithmic terms $\cO{\asb^n\ga^{-k}}$ for $2n+1 -N \le k \le
2n+1$, and $n >N$. 

Hence if we can construct a kernel which has no double logarithms,
$\as^n\ga^{-(n+1+k)}$, $k>0$, for a scale $s_0=k_1^2$, no DLs,
$\as^n(1-\ga)^{-(n+1+k)}$, for a scale $s_0=k_2^2$, and which is exact
to order $N$, we will ``for free'' correctly obtain, for scale
$s_0=k_1k_2$ all DLs, $\asb^n\ga^{-k}$ and $\asb^n(1-\ga)^{-k}$ with
$2n+1 -N \le k \le 2n+1$. More specifically, for the NLL case the aim
is to construct a kernel whose most divergent terms for $s_0=k_1^2$
are $\as^n/\ga^{n+1}$, and analogously for scale $s_0=k_2^2$ ---
modifying the most divergent unknown term $\as^2/\ga^3$ can then at
worst change the sub-sub-leading double logarithms,
$\as^n/\ga^{2n-1}$, $n\ge2$. In other words, the leading and
sub-leading double logarithms will be known unambiguously

So returning to the general case, let our resummed kernel be denoted
by $\chi^{(N)}(\ga)$ (for scale $s_0=k_1k_2$), with a perturbative
expansion
\begin{equation*}
  \chi^{(N)}(\ga) = \sum_{n=0}^N \asb^n \chi_n(\ga) + 
  \sum_{n=N+1}^\infty \asb^n \chi^{(N)}_n(\ga)\,,
\end{equation*}
where $\chi_n$ is the exact order $n$ contribution to the BFKL kernel.

For each order $n\ge0$ of the exact kernel, $\chi_n(\ga)$, there will
be divergences $d_{n,k}/\gamma^k$, for $1\le k\le 2n+1$, with
$d_{n,k}$ being numerical coefficients. There will be similar
divergences at $\gamma=1$ with coefficients ${\bar d}_{n,k}$.
Furthermore, $\chi^{(N)}_n(\ga)$ will have divergences
$\asb^n/\gamma^k$ with coefficients $d^{(N)}_{n,k}$ (and similarly for
$(1-\ga)$). These coefficients will have the property
\begin{equation*}
  d^{(N)}_{n,k} = d_{n,k} \quad \text{ for } \quad (2n+1 -N \le k \le
  2n+1) \,.
\end{equation*}
To construct a kernel $\chi^{(N)}$, it is useful to specify a set of
functions $D_k(\gamma)$ which are regular for $\gamma>1/2$, and at
$\gamma=0$ have only the divergence $1/\gamma^k$. We can then
immediately construct a kernel $\chi^{(0)}(\ga)$ which is free of DLs
in the appropriate limits and which is exact to $\cO{\asb^0}$:
\begin{multline}
  \label{eq:chi0}
  \chi^{(0)}(\ga) = \chi^{(0)}(\ga,\omega=\asb\chi^{(0)}) =\\
  \chi_0(\ga) +d_{0,1} 
  [D_1(\ga+\half\omega) - D_1(\ga)]
  +{\bar d}_{0,1}
  [D_1(1-\ga+\half\omega) - D_1(1-\ga)]\,.
\end{multline}
To go to $N>0$ one uses the recursion relation
\begin{multline}
  \label{eq:chiNrec}
  \chi^{(N)}(\ga) =
  \chi^{(N)}\left(\ga,\omega=\asb\chi^{(N)}\right) =
  \chi^{(N-1)}(\ga,\omega)+ \\ + \asb^N\left(\chi_N(\ga) -
    \chi^{(N-1)}_N(\ga)\right) \,+\,
  \asb^N \sum_{k=1}^{N+1} \left[\left(d_{N,k} -  d_{N,k}^{(N-1)}\right)
  [D_k(\ga+\half\omega) - D_k(\ga)] \right. \\ \left.
  +\left({\bar d}_{N,k}-{\bar d}_{N,k}^{(N-1)}\right)
  [D_k(1-\ga+\half\omega) - D_k(1-\ga)]\right].
\end{multline}
To demonstrate that $\chi^{(N)}$ correctly reproduces the kernel up to 
order $\asb^N$ it suffices to note that neither the parts in square
brackets, nor 
\begin{equation*}
  \chi^{(N-1)}(\ga,\omega) - \asb^N\chi^{(N-1)}_N(\ga)
\end{equation*}
contains any terms at $\cO{\asb^N}$, so that the entire $\cO{\asb^N}$
part comes from $\chi_N(\ga)$. 

The next step is to show that the characteristic function written in
terms of the scale $s_0=k_1^2$, $\xi(\ga) = \chi(\ga-\ho)$ is free of
DL divergences, $\asb^n\ga^{-k}$ for $k>n+1$. It is convenient to
rewrite the kernel as:
\begin{multline}
  \label{eq:chiN}
  \chi^{(N)}(\ga) =
  \chi^{(N)}\!\left(\ga,\omega=\asb\chi^{(N)}\right) = 
  \sum_{n=0}^N \asb^n R_n(\ga) + \\
  +\sum_{n=0}^N \asb^n \sum_{k=1}^{n+1} \left[
     \left(d_{n,k} -  d_{n,k}^{(n-1)}\right)D_k(\ga+\ho) 
     +\left(\bd_{n,k} -  \bd_{n,k}^{(n-1)}\right)D_k(1-\ga+\ho) 
    \right],
\end{multline}
where
\begin{equation}
  \label{eq:Rn}
  R_n(\ga) =   \chi_n(\ga)  -
    \chi^{(n-1)}_n(\ga) - \sum_{k=1}^{n+1} \left[
     \left(d_{n,k} -  d_{n,k}^{(n-1)}\right)D_k(\ga) 
     +\left(\bd_{n,k} -  \bd_{n,k}^{(n-1)}\right)D_k(1-\ga) 
    \right],
\end{equation}
with $\chi^{(-1)}=0$ and all $d^{(-1)}_{n,k}=0$. The fact that
$\chi^{(n-1)}_n(\ga)$ has the same pattern of $\ga^{-k}$ and
$(1-\ga)^{-k}$, $k>n+1$, divergences as $\chi_n(\ga)$ means that
$R_n(\ga)$ is regular, for each $n$, at $\ga=0$ and $\ga=1$.
Applying the transformation $\ga\to\ga-\ho$ to get $\xi(\ga)$ gives us
\begin{multline}
  \label{eq:xiN}
  \xi^{(N)}(\ga) =
  \xi^{(N)}\!\left(\ga,\omega=\asb\,\xi^{(N)}\right) = 
  \sum_{n=0}^N \asb^n R_n(\ga-\ho) + \\
  +\sum_{n=0}^N \asb^n \sum_{k=1}^{n+1} \left[
     \left(d_{n,k} -  d_{n,k}^{(n-1)}\right)D_k(\ga) 
     +\left(\bd_{n,k} -  \bd_{n,k}^{(n-1)}\right)D_k(1-\ga+\om) 
    \right].
\end{multline}
The first half of the second line produces a set of divergences
$\asb^n\ga^{-k}$, with $k \le n+1$; i.e.\ that part is free of double
logarithms. Substituting these into the second half of the second line
and into $R_n$ will at worst produce extra divergences
$\asb^n\ga^{-n}$. Hence $\xi(\ga)$ is free of double logarithmic
divergences, and exact to $\cO{\asb^n}$ (since $\chi^{(N)}$ was exact
to $\cO{\asb^n}$). Similarly changing scale to $s_0=k_2^2$, there will
be no DL divergences at $\ga=1$.

This completes the demonstration that $\chi^{(N)}$ is exact up to
$\cO{\asb^n}$ and that it correctly reproduces the double logarithms
$\asb^n\gamma^{-k}$ and $\asb^n(1-\gamma)^{-k}$ with $2n+1-N \le k\le
2n+1$.

Analytically the expression for the series of the leading double
logarithms has been given in (\ref{eq:chiLser},
\ref{eq:chiLsercoeffs}). The sub-leading double logarithms,
$\asb^n/\ga^{2n}$, are determined entirely by the coefficient of
$\asb/\ga^2$, $d_{1,2}$ (which itself depends on one's choice of scale
for $\as$):
\begin{equation*}
  d_{n,2n} = (-1)^{n+1}\, d_{1,2}\, \frac{(2n-1)!}{n!\,(n-1)!\, 2^{n-1}}\,.
\end{equation*}


\subsection{Resummation schemes}
\label{sec:resumsc}

The procedure described above is not unique in that there is some
freedom in the choice of the divergent functions $D_k(\ga)$. This
freedom results in differences in the non-guaranteed terms of
$\chi^{(N)}$, and is in some sense a measure of the uncertainties due
to unknown higher order terms.

To study the properties of the resummed NLL kernel, and the
ambiguities which remain, we will consider four different choices for
the treatment of uncontrolled subleading terms. Extra details
(including the explicit forms for schemes 1 and 3) are to be found in
the appendix.
\begin{enumerate}
\item A choice similar to that used in the toy model:
  \begin{equation*}
    D_k(\gamma) = \frac{(-1)^{k-1}}{(k-1)!}
    \frac{d^{k-1}}{d\gamma^{k-1}} \left[ 
      \psi(1) - \psi(\gamma)
    \right]\,.
  \end{equation*}
\item Even simpler, is
  \begin{equation*}
    D_k(\gamma) = \frac{1}{\gamma^k}\,.
  \end{equation*}
\item Both of these choices will run into problems at some value of
  $\as$, because even after resumming double-logarithmic contributions
  there are terms such as $\asb^n/\gamma^{n+1}$ which can be dangerous.
  Their resummation is related to the reproduction of the correct
  DGLAP limit and a resummation of the running of $\asb$. Both should
  not be difficult to do, but for simplicity will not be done exactly.
  We will simply modify the resummation procedure suggested above and
  for $\chi^{(0)}$ use
  \begin{multline}
    \chi^{(0)}(\gamma) = \chi^{(0)}(\gamma,\omega=\asb\chi^{(0)}) 
    = \\ (1-\asb A) \left[2\psi(1) - \psi(\gamma+\ho+\asb B) 
      -\psi(1-\gamma+\ho+\asb B) \right]\,,
  \end{multline}
  with the constants $A$ and $B$ chosen such that $d^{(0)}_{1,k} =
  d_{1,k}$ for $k\ge 1$, not just for $k=3$. The choice of the $D_k$ is
  then immaterial since they are not needed at $\cO{\as}$.
\item A similar procedure (but more in line with scheme 2) is to use
  \begin{multline}
    \chi^{(0)}(\gamma) = \chi^{(0)}(\gamma,\omega=\asb\chi^{(0)}) 
    = \chi_0(\ga) - \frac{1}{\ga} - \frac{1}{1-\ga} +\\
    + (1-\asb A) \left[\frac{1}{\gamma+\ho+\asb B} +
      \frac{1}{1-\gamma+\ho+\asb B} 
       \right]\,,
  \end{multline}
  where again $A$ and $B$ are chosen such that $d^{(0)}_{1,k} = d_{1,k}$
  for $k\ge 1$.
\end{enumerate}
\begin{figure}[htbp]
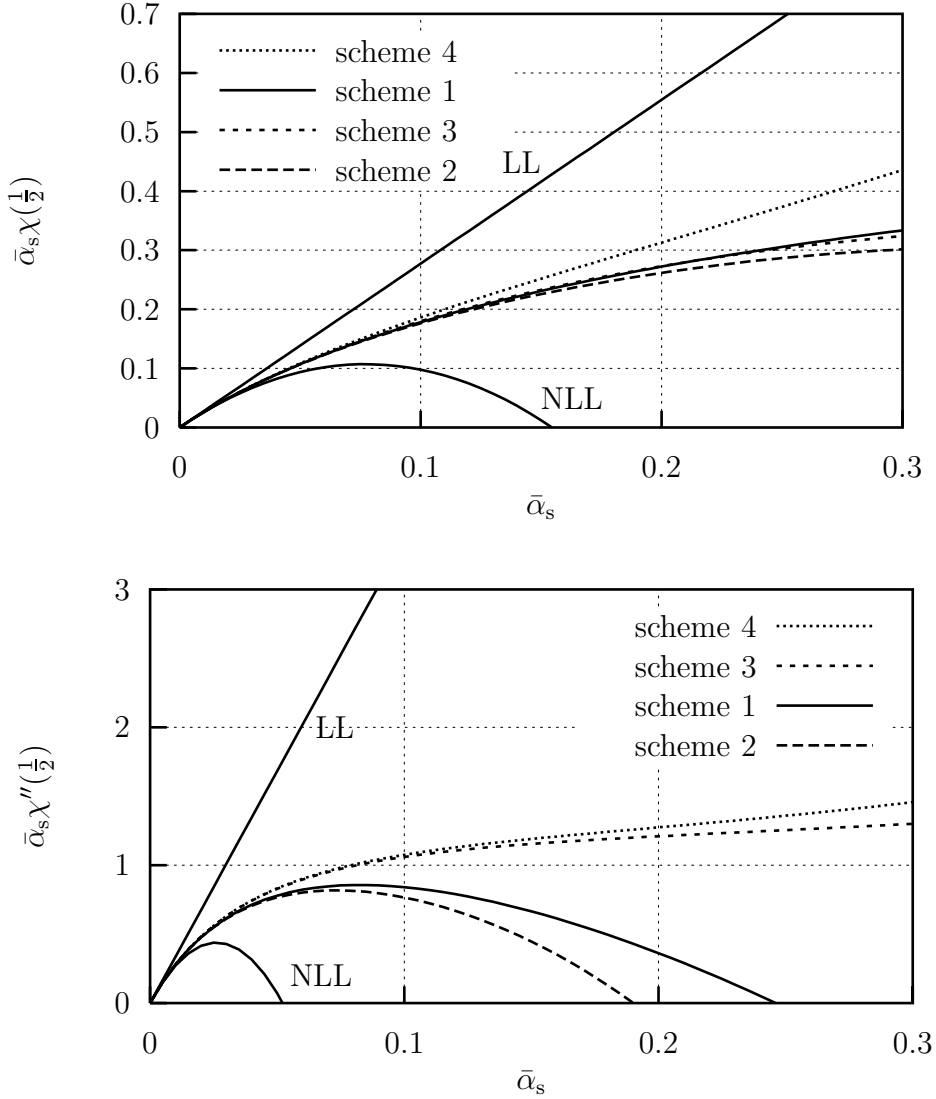

  \begin{center}
    \input{ltnl.pstex}
    \input{ltnl.deriv.pstex}
    \caption[]{The result of resummation on $\chi$ and its second
      derivative; shown for $\nf=0$. A symmetric version of the pure
      NLL kernel has been used, corresponding to the suggestion
      \cite{FL} to redefine $k^{2(\ga-1)}$ as
        $\sqrt{\as(k^2)}k^{2(\ga-1)}$ in the eigenfunctions.}
    \label{fig:chires}
  \end{center}
\end{figure}
Results using these four resummation schemes are shown in
figure~\ref{fig:chires}. The value of $\asb\chi(\half)$ turns out to
be relatively stable under changes of scheme --- and quite different
from the pure NLL result. Indeed it seems somewhat more in line with
what one might expect from phenomenology. The reason for the larger
difference between scheme 4 and the other schemes is not entirely
clear.

The results for $\asb\chi''(\half)$ show quite a considerable spread.
This is connected with the treatment of the $\asb/\ga^2$ and
$\asb/(1-\ga)^2$ terms, which remain ``untamed'' even after the
resummation of the $\as^n\ga^{-(2n+1)}$ and $\as^n\ga^{-2n}$ terms. In
practice they have quite a large coefficient and precisely because
they are $\propto 1/\ga^2$ their contribution to the second derivative
is enhanced (by a factor of $3$ naively) as compared to their
contribution to $\chi(\half)$. Once the $\as^n\ga^{-(n+1)}$ terms are
resummed (which fixes also the DL terms $\as^n\ga^{-(2n-1)}$), as in
schemes 3 and 4 (which resum $\as^n\ga^{-(n+1)}$ terms in the same way),
one finds a certain independence on the exact resummation scheme used
for the DLs. It should be emphasised however that the scheme used here
for resumming the $\as^n\ga^{-n}$ terms has been chosen {\it ad hoc},
so that all that should be taken seriously is the qualitative effect
of a sufficient resummation, namely that it ensures that the second
derivative stays positive, and that the saddle point at $\ga=\half$
does not split into two.

\section{Discussion}
\label{sec:discuss}

The main message of this paper is that the pathological aspects of the
NLL BFKL kernel have to be, and can be cured by suitable
resummations.\footnote{This differs from the conclusions presented in
  \cite{BV}, where it is suggested that the calculation of the NNLL
  and NNNLL-order terms would suffice to improve the convergence. The
  origin of the difference in conclusions lies perhaps in the fact
  that the authors of \cite{BV} study predominantly the anomalous
  dimension (the expansion around $\ga=0$, which is less sensitive to
  double-logarithms from $\ga=1$) while the aim of this paper is to
  examine the properties of the kernel around $\ga=1/2$.}  A first
step along the way is to resum leading and sub-leading double
logarithms, which modifies the NLL kernel as shown in
figure~\ref{fig:chires}.

The approach used here for determining the double logarithms relies on
two elements: the absence of double transverse logarithms in the DGLAP
limits (with scale choices $k_1^2$ or $k_2^2$) and the assumption that
one can change the scale of the kernel consistently at all orders by a
transformation of the form
\begin{equation*}
  \label{eq:scalechange}
  \ga \to \ga \pm \half \asb\chi(\ga)\,.
\end{equation*}
This is tied in with the assumption that the kernel $\asb\chi(\ga)$
exponentiates, i.e.\ that one can write the gluon's Green function as
in \eqref{eq:eigen}. At NNLL order, one knows that exponentiation no
longer holds, due for example to the running of $\as$
\cite{KoMu,Levin} and unitarity corrections, so that a change of scale
is no longer accomplished by a simple shift of $\ga$. To try to
understand the effect of changing the scale in a kernel which fails to
exponentiate at NNLL order, one can take as a rough example, inspired
by \cite{KoMu,Levin}, a worst-case situation where the
non-exponentiation results in
\begin{equation*}
  \exp\left(\frac{\as\ln s}{\ga}\right) \to 
  \exp\left(\frac{\as\ln s}{\ga} + \cO{\frac{\as^5 \ln^3
        s}{\ga^5}}\right).
\end{equation*}
Working through the effect of a change in scale of $s$, one finds that
the shift in $\ga$ breaks down at a level which can affect terms
$\as^n/\ga^{2n-1}$ in the $s_0=k_1k_2$ kernel. This suggests that the
leading and sub-leading DLs worked out in this paper should not be
affected by the non-exponentiation of the kernel.


The resummation of double logarithms is probably sufficient to obtain
a reasonable idea of the typical power that one may expect at small
$x$.  But to carry out serious phenomenology it is vital to have
adequate control over the value of the second derivative,
$\chi''(\half)$. This requires the additional resummation of
$\as^n\ga^{-(n+1)}$ terms. A qualitative illustration of the effect of
such a resummation is given with schemes 3 and 4 in
figure~\ref{fig:chires}. A correct treatment would involve a
resummation of collinear logarithms (which perhaps can be accomplished
\cite{CC97a} simply by shifting the $1/\ga$ to $1/(\ga+\cO{\as})$ as
done in schemes 3 and 4) and additionally the resummation of
logarithms associated with the running of $\as$, which will behave
differently: for example evaluating the kernel with $\as$-scale $k_1$,
close to $\ga=1$, accounting just for running coupling effects in a
single emission, one will have
\begin{align*}
  \as(k_1)\chi(\ga) &\sim \asb(k_1^2) \int_{k_1^2}^\infty
  \frac{dk_2^2}{k_2^2} 
   \left(\frac{k_2^2}{k_1^2}\right)^{\ga-1} \frac{1}{1 + \asb(k_1^2)\,
     \beta_0  \ln 
     k_2^2/k_1^2} \\
   &= \asb(k_1^2)\sum_{n=0}^\infty (-1)^n n!
   \frac{(\asb\beta_0)^n}{(1-\ga)^{n+1}}\,. 
\end{align*}
which sums to an incomplete $\Gamma$-function, and has a distinctly
different structure from the resummation of collinear logarithms ---
so there remains the job of putting them together. There are of course
many other issues associated with the running of $\as$. These have
already been considered in some detail by a number of groups
\cite{KoMu,Levin,CC96}. Finally it might be possible to understand, at
all orders, contributions which are enhanced neither at $\ga=0$ nor at
$\ga=1$ but which are still important at $\ga=\half$, such as those
arising from angular ordering \cite{CCFM,BMSS97}.

\acknowledgments 

I would like to acknowledge a stimulating conversation with Victor
Fadin about the poor convergence of the kernel and the presence of
double logarithms.  I am grateful also to Johannes Bl\"umlein,
Marcello Ciafaloni, Yuri Dokshitzer, Gosta Gustafson, Giuseppe
Marchesini and Douglas Ross for helpful discussions.

\appendix

\section{Application of the resummation to the NLL kernel}
\label{sec:appnll}

For completeness, it is useful to discuss the application to the NLL
kernel of the resummation procedure discussed in
section~\ref{sec:resum}. It will be illustrated explicitly for the
cases of schemes 1 and 3, using a symmetric version (corresponding to
the extraction of $\asb(k_1k_2)$, or
$\sqrt{\asb(k_1^2)\asb(k_2^2)}\,$) of the NLL kernel:
\begin{multline}
  4\chi_1(\ga) = 
  -\left[  \frac{2\beta_0}{\ca} \,\chi_0^2(\ga ) - 
    K \chi_0 (\ga )
 -6\zeta (3)
\right . \\ \left.
+ \frac{\pi ^2\cos(\pi \ga )}{\sin^2(\pi \ga
)(1-2\ga )}\left( 3+\left( 1+\frac{n_f}{\ca^3}\right) \frac{2+3\ga
(1-\ga )}{(3-2\ga )(1+2\ga )}\right) 
\right. \\ \left.
 -\psi ^{\prime \prime }(\ga )-\psi ^{\prime \prime }(1-\ga )-%
\frac{\pi ^3}{\sin (\pi \ga )}+4\phi (\ga )\right] \,,
\end{multline}
where
\begin{align*}
  \beta_0 &= \frac{11\ca}{12} - \frac{2\nf}{12}\,,\\
  K &= \frac{67}9 - \frac{\pi^2}{3} - \frac{10\nf}{9\ca}\,,
\end{align*}
and
$$
\phi (\gamma )=\sum_{n=0}^\infty (-1)^n\left[ \frac{\psi (n+1+\gamma
    )-\psi (1)}{(n+\gamma )^2}+\frac{\psi (n+2-\gamma )-\psi
    (1)}{(n+1-\gamma )^2}\right] \,. 
$$
First we must determine the pattern of divergences $\as^n/\ga^k$
(${\bar d}_{n,k}=d_{n,k}$) of the LL and NLL kernels:\footnote{One
  notes that $\nf$-independent part of $d_{1,1}$ is zero.}
\begin{subequations}
\begin{align*}
  d_{0,1} &= 1 \,,\\
  d_{1,1} &= -\frac{67}{36} - \frac{13\nf}{36\ca^3} +
  \frac{\pi^2}{12} + \frac{K}{4}\,,
  \\
  d_{1,2} &= -\frac{\beta_0}{2\ca} - \frac{11}{12} - \frac{\nf}{6\ca^3}\,,
  \\
  d_{1,3} &= -\frac12 \,.
\end{align*}
\end{subequations}
The steps needed to resum the DLs in the kernel then depend on the
scheme used.

\subsection{Scheme 1}
\label{app:schm1}
In scheme 1, the terms used to subtract off the divergences $1/\ga^k$,
$D_k(\ga)$ are
\begin{subequations}
\begin{align*}
  D_1(\ga) &= \psi(1)-\psi(\ga)\,,\\
  D_2(\ga) &= \psi'(\ga)\,.
\end{align*}
\end{subequations}
Applying \eqref{eq:chi0}, the kernel correct to LL order and with all
the leading DLs is just the toy-kernel discussed in
section~\ref{sec:ldc}:
\begin{equation*}
  \chi^{(0)}(\ga) = \chi^{(0)}(\ga,\om=\asb\chi^{(0)}) = 
2\psi(1) - \psi(\ga + \half\omega) -
  \psi(1-\ga+\half\omega)\,.
\end{equation*}
The next step is to determine the NLL part of $\chi^{(0)}$ (to avoid
any double-counting when one adds in the pure NLL part of the kernel).
It is given by \eqref{eq:chi1}:
\begin{equation*}
  \chi^{(0)}_1 = -\frac12\chi_0(\ga)[\psi'(\ga) + \psi'(1-\ga)]\,,
\end{equation*}
and the coefficients of its divergences (${\bar
  d}_{n,k}^{(0)}=d_{n,k}^{(0)}$) are
\begin{align*}
  d_{1,1}^{(0)} &= -\frac{\pi^2}{6}\,, \\
  d_{1,2}^{(0)} &= 0\,, \\
  d_{1,3}^{(0)} &= -\frac12\,.
\end{align*}
The fact that $d_{1,3}^{(0)}=d_{1,3}$ is a consequence of $\chi^{(0)}$
correctly reproducing the DLs that are present at NLL order. Using
\eqref{eq:chiNrec} we are now in a position to write down
$\chi^{(1)}$, i.e.\ the kernel which is correct to NLL order and
contains the appropriate leading and sub-leading DLs.  Essentially it
is obtained by ``shifting'' the divergences which are not already
accounted for by $\chi^{(0)}_1$
\begin{multline}
\label{eq:chi1summed}
  \chi^{(1)}(\ga) = 
  \chi^{(0)}\left(\ga,\om=\asb\chi^{(1)}\right) = 
  \chi^{(0)}(\ga,\om) + \asb\left(\chi_1(\ga) - \chi_1^{(0)}\right)\\
  + \sum_{k=1}^2
  \asb(d_{1,k} - d^{(0)}_{1,k})\left( D_k(\ga+\ho)+D_k(1-\ga+\ho)
    -D_k(\ga) - D_k(1-\ga)\right)\,.
\end{multline}
This equation applies also to scheme 2, as long as one modifies
$\chi^{(0)}$, $\chi^{(0)}_1$ and $d_{1,k}^{(0)}$ appropriately.

\subsection{Scheme 3}
\label{app:schm3}
Schemes 3 and 4 differ from schemes 1 and 2 in the form used for
$\chi^{(0)}$, which is constructed in such a way that
$d_{1,k}^{(0)}=d_{1,k}$ not just for $k=3$, but also for $k=1,2$; they
resum (in a manner which given the present state of the art is
arbitrary) the $\as^n/\ga^n$ and $\as^n/\ga^{n+1}$ divergences.  For
scheme 3, where one uses
\begin{multline}
  \label{eq:chi0sch3}
  \chi^{(0)}(\gamma) = \chi^{(0)}(\gamma,\omega=\asb\chi^{(0)}) 
  = \\ (1-\asb A) \left[2\psi(1) - \psi(\gamma+\ho+\asb B) 
    -\psi(1-\gamma+\ho+\asb B) \right]\,,
\end{multline}
the NLL part of $\chi^{(0)}$ is 
\begin{equation*}
  \chi^{(0)}_1 = -\left(B + \frac12\chi_0(\ga)\right)
  [\psi'(\ga) + \psi'(1-\ga)] - A \,\chi_0\,,
\end{equation*}
and its divergences are given by
\begin{subequations}
\begin{align*}
  d_{1,1}^{(0)} &= -A-\frac{\pi^2}{6}\,, \\
  d_{1,2}^{(0)} &= -B\,, \\
  d_{1,3}^{(0)} &= -\frac12\,.
\end{align*}
\end{subequations}
Since we want $\chi^{(0)}_1$ to reproduce the divergences of $\chi_1$, 
i.e.\ $d_{1,k}^{(0)}=d_{1,k}$,
we have to set $A$ and $B$ as
\begin{subequations}
\begin{align*}
  A &= d_{1,1} - \frac{\pi^2}{6}\,,\\
  B &= d_{1,2}\,.
\end{align*}
\end{subequations}
One notes (by carrying out the shift) that $\chi^{(0)}$ is free of the
appropriate double logarithms when one changes scale to $k_1^2$ or
$k_2^2$. To obtain the version of the kernel which is correct to NLL
order, one then applies a similar procedure to that of
\eqref{eq:chiNrec}, except that there is no need for any piece
analogous to the second line of \eqref{eq:chi1summed}, since
$d_{1,k}^{(0)}=d_{1,k}$:
\begin{equation}
  \chi^{(1)}(\ga) = 
  \chi^{(0)}\left(\ga,\om=\asb\chi^{(1)}\right) = 
  \chi^{(0)}(\ga,\om) + \asb\left(\chi_1(\ga) - \chi_1^{(0)(\ga)}\right)\,.
\end{equation}
Since the second term on the RHS is finite at $\ga=0$ and $\ga=1$, the
fact that $\chi^{(0)}$ is free of the relevant double logarithms when
transformed to scales $k_1^2$ or $k_2^2$, is sufficient to guarantee
that $\chi^{(1)}$ has this property too, and hence that for scale
$k_1k_2$ it correctly reproduces the leading and sub-leading DLs.

The procedure used for scheme 4 is analogous.


\begin{thebibliography}{99}
\bibitem{NLL} 
  L. N. Lipatov and V. S. Fadin, \sjnp{50}{1989}{712}; \\
  V. S. Fadin, R. Fiore and M. I. Kotsky, \plb{539}{1995}{181}; \\ 
  V. S. Fadin, R. Fiore and M. I. Kotsky, \plb{387}{1996}{593}
  [\hepph{99605357}]; \\ 
  V. S. Fadin, and L. N. Lipatov, \npb{406}{1993}{259}; \\ V. S. Fadin,
  R. Fiore and A. Quartarolo, \prd{50}{1994}{5893} [\hepth{9405127}];
  \\ V. S. Fadin, 
  R. Fiore, and M. I. Kotsky, \plb{389}{1996}{737} [\hepph{9608229}];\\
  V. S. Fadin and L. N. Lipatov, \npb{477}{1996}{767} [\hepph{9602287}]; \\
  V. S. Fadin, M. I. Kotsky and L. N. Lipatov, \plb{415}{1997}{97}; \\
  S. Catani, M. Ciafaloni and F.Hautman, \plb{242}{1990}{97}; \\
  S. Catani, M. Ciafaloni and F.Hautman, \npb{366}{1991}{135}; \\ 
  V. S. Fadin, R. Fiore, A. Flashi, and M.
  I. Kotsky, \plb{422}{1998}{287} [\hepph{9711427}].


\bibitem{BFKL} L.N. Lipatov, Sov. J. Phys. 23 (1976) 338;\\
       E.A. Kuraev, L.N. Lipatov and V.S. Fadin, Sov. Phys. JETP 45
       (1977) 199; \\
       Ya. Balitskii and L.N. Lipatov, Sov. J. Nucl. Phys. 28 (1978)
       822.
\bibitem{FL} V.S. Fadin and L.N. Lipatov, \hepph{9802290}.
\bibitem{CC98}  M. Ciafaloni and G. Camici, \hepph{9803389};\\
  M. Ciafaloni, \plb{429}{1998}{363} [\hepph{9801322}];\\
  M. Ciafaloni and G. Camici, \plb{412}{1997}{396} [\hepph{9707390}]. 
\bibitem{BV} J. Bl\"umlein and A. Vogt, \prd{57}{1998}{1}
  [\hepph{9707488}];\\ 
 J. Bl\"umlein and A. Vogt, \prd{58}{1998}{014020} [\hepph{9712546}];\\
 J. Bl\"umlein, W.L. van Neerven, V. Ravindran, A. Vogt,
    \hepph{9806368};\\
    J. Bl\"umlein, W.L. van Neerven, V. Ravindran,
    \hepph{9806357}.

\bibitem{Ross} D.A. Ross, \plb{431}{1998}{161} [\hepph{9804332}].
\bibitem{Levin} E. Levin, \hepph{9806228}.
\bibitem{DGLAP} V.N. Gribov and L.N. Lipatov, \sjnp{15}{1972}{438};\\
  G. Altarelli and G. Parisi, \npb{126}{1977}{298}; \\ Yu.L. Dokshitzer,
  \jetp{46}{1977}{641}. 
\bibitem{LDC} B. Andersson, G. Gustafson, J. Samuelsson,
  \npb{467}{1996}{443}. 
\bibitem{LDCmc} H. Kharraziha and L. L\"onnblad, \jhep{03}{1998}{006}
  [\hepph{9709424}]. 
\bibitem{future} G. Bottazzi, G. Marchesini, G.P. Salam and
  M. Scorletti, in preparation.
\bibitem{CCFM}
  M. Ciafaloni, \npb{296}{1987}{249};\\
  S. Catani, F. Fiorani and G. Marchesini, \plb{234}{1990}{339};\\
  S. Catani, F. Fiorani and G. Marchesini, \npb{336}{1990}{18}.
\bibitem{KoMu} Yu.V. Kovchegov and A.H. Mueller, \hepph{9805208}.
\bibitem{CC97a} M. Ciafaloni and G. Camici, \npb{496}{1997}{305}
  [\hepph{9701303}].  
\bibitem{CC96} G. Camici and M. Ciafaloni, \plb{395}{1997}{118}
  [\hepph{9612235}].
\bibitem{BMSS97} G. Bottazzi, G. Marchesini, G.P. Salam and
  M. Scorletti, \npb{505}{1997}{366} [\hepph{9702418}].


\end{thebibliography}
\end{document}